\documentclass{aanew}
\usepackage{psfig}
\usepackage{graphicx}

\def\bron{4U~1254-690}
\def\ecs{erg~cm$^{-2}$s$^{-1}$}
\def\lum{erg~s$^{-1}$}
\renewcommand{\tabcolsep}{1.5mm} 

\begin{document}

\title{A superburst from \bron}

\titlerunning{A superburst from \bron} 
\authorrunning{J.J.M. in 't Zand et al.}

\author{
J.J.M.~in~'t~Zand\inst{1,2},
E.~Kuulkers\inst{3},
F.~Verbunt\inst{2},
J.~Heise\inst{1,2},
R.~Cornelisse\inst{4}
}

% \offprints{J.J.M. in 't Zand, email {\tt jeanz@sron.nl}}

\institute{     SRON National Institute for Space Research, Sorbonnelaan 2,
                3584 CA Utrecht, the Netherlands 
	 \and
                Astronomical Institute, Utrecht University, P.O. Box 80000,
                3508 TA Utrecht, the Netherlands
	 \and
                ESA-ESTEC, Science Ops. \& Data Systems Div.,
                SCI-SDG, Keplerlaan 1, 2201 AZ Noordwijk, the Netherlands
         \and
                Dept. of Physics and Astronomy, University of Southampton,
                Hampshire SO17 1BJ, U.K.
	}

\date{Received, accepted }

\abstract{We report the detection with the BeppoSAX Wide Field Cameras
of a superburst from \bron. The superburst is preceded by a normal
type-I X-ray burst, has a decay time that is the longest of all eight
superbursts detected so far and a peak luminosity that is the
lowest. Like for the other seven superbursts, the origin is a
well-known type-I X-ray burster with a persistent luminosity level
close to one tenth of the Eddington limit. Based on WFC data of all
persistently bright X-ray bursters, the average rate of superbursts is
$0.51\pm0.25$ per year per persistently bright X-ray burster. Some
systems may have higher superburst rates. For all superbursters, we
present evidence for a pure helium layer which is burnt in an unstable
{\em as well as a stable} manner.  }

\maketitle 

\section{Introduction}
\label{intro}

\bron\ is a persistently bright low-mass X-ray binary (Griffiths et
al. 1978) exhibiting type-I X-ray bursts (Mason et al. 1980;
Courvoisier et al. 1986). A type-I X-ray burst is a seconds-to-minutes
long X-ray event caused by a thermonuclear flash in the upper layers
of a neutron star that is receiving matter from a nearby, usually
Roche-lobe filling, companion star (for reviews see Lewin et al. 1993
and Strohmayer \& Bildsten 2003). Courvoisier et al. (1986) discovered
periodic dipping activity with a period of $3.88\pm0.15$~hr and
identified this as the binary orbital period. The dips are caused by
obscuration of the central source by a bulge on the outer edge of the
accretion disk. The inclination is close to edge-on but not completely
because no eclipses are observed (68$<i<$73; Motch et
al. 1987). Motch et al. and Courvoisier et al. estimate through
independent methods distances of 9--14 and 8--15 kpc
respectively. Recent X-ray studies by Smale et al. (1999; 2002), Iaria
et al. (2001) and Boirin \& Parmar (2003) show that dipping activity
is not present during each orbit.

A few years ago the BeppoSAX Wide Field Cameras (WFCs; Jager et
al. 1997) brought about the discovery of thermonuclear flashes with
durations and energetics 10$^3$ times higher than for ordinary type-I
X-ray bursts (Cornelisse et al. 2000). These so-called superbursts are
rare, with an estimated frequency of once per few years per object. So
far, seven superbursts have been detected from six sources: one from
4U~1735-44 (Cornelisse et al. 2000), Ser X-1 (Cornelisse et al. 2002),
KS~1731-260 (Kuulkers et al. 2002), 4U~1820-303 (Strohmayer \& Brown
2002), GX 3+1 (Kuulkers 2002) and two from 4U~1636-536 within 4.7~yr
(Wijnands 2001; Strohmayer \& Markwardt 2002). It is thought that
unstable carbon burning (Woosley \& Taam 1976; Strohmayer \& Brown
2002) in a heavy-element ocean (Cumming \& Bildsten 2001), possibly
combined with photo-disintegration-triggered nuclear energy
release (Schatz et al. 2003), is responsible for most superbursts.

In the present paper we report the WFC detection of the eighth
superburst, from \bron.

\section{Burst detections}
\label{obs}

\begin{figure}[t]
\psfig{figure=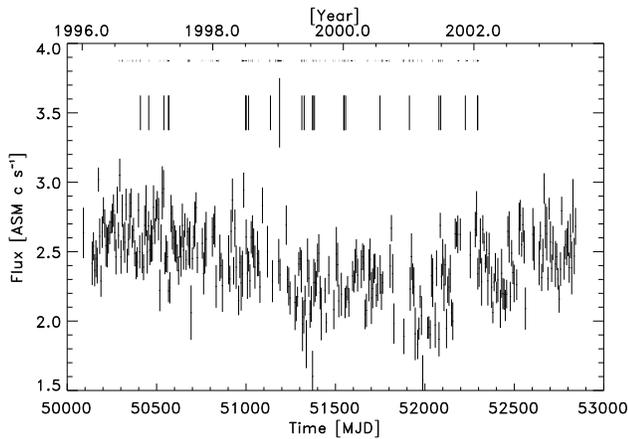,width=\columnwidth,clip=t}
\caption{2--12 keV RXTE-ASM light curve of \bron\ at a 1-week time
resolution. Data points with errors in excess of 0.2~c~s$^{-1}$ were
excluded from this plot. The horizontal lines in the top indicate the
WFC coverage of the source and the vertical lines the bursts detected
with the WFC.  The long vertical line indicates the time of the
superburst.
\label{figasm}}
\end{figure}

Measurements with the RXTE All-Sky Monitor (ASM; Levine et al. 1996)
provide the most complete picture of the persistent 2-12 keV flux from
\bron\ in recent years, see Fig.~\ref{figasm}. They reveal that the
source is persistently present. There are no strong week-to-week
variations, but there is a slow trend: during 1999 through 2000 it was
about 30\% fainter than before or after. The average flux is
$2.42\pm0.01$ ASM c~s$^{-1}$, or 32~mCrab. This is, within a few tens
of percents, equal to many previous measurements (Courvoisier et
al. 1986; Uno et al. 1997; Iaria et al. 2001; Boirin \& Parmar 2003),
although in the 1970s \bron\ appears to have been much more variable
on long time scales (see Griffiths et al. 1978 and references
therein).

The BeppoSAX WFCs acquired a total exposure time of 10.3~Ms on \bron.
This exposure includes times when the satellite attitude
reconstruction is not optimum. The detrimental effect on the quality
of the data is only limited when considering {\em brief} X-ray bursts
because the attitude is not expected to change much within a few tens
of seconds. Confinement to good-attitude data would result in ignoring
28\% of the data.

\begin{figure}[t]
\psfig{figure=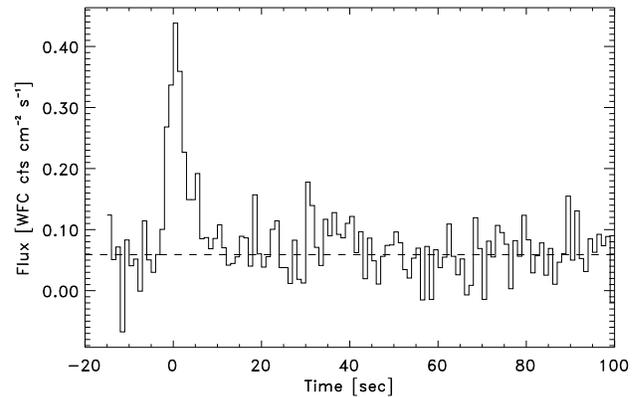,width=\columnwidth,clip=t}
\caption{The average 2--28 keV profile over all normal bursts
observed from \bron, except the one related to the superburst.
The dashed line refers to the persistent flux level.
\label{figavburst}}
\end{figure}

Unfortunately, the source is rather distant. This implies that X-ray
bursts, whose luminosity can at maximum peak at the Eddington limit,
are relatively faint. For 16~kpc the 2--10 keV peak flux would be of
order one third that of the Crab. If the bursts are short, as appears
to be the case for the six X-ray bursts reported so far in some detail
(Courvoisier et al. 1986; Smale et al. 2002), they are near the
detection limit of the WFCs. Nevertheless, we carried out a systematic
search for bursts. To optimize the detection limit for expected
bursts, we searched with a time resolution of 4 sec in the full
bandpass for increases above the persistent flux level. The 4$\sigma$
detection threshold varies between 0.2 and 0.3 Crab units (2--28
keV). In total 26 bursts were identified.  Their times are indicated
in Fig.~\ref{figasm}. The shortest wait time between two bursts is
2.5~hr and sub-day wait times were observed in 4 other pairs of bursts
(see also Smale et al. 2002). The burst peak flux varies between 0.4
and 0.7~WFC~c~s$^{-1}$cm$^{-2}$ (0.2 to 0.3 Crab units). Most bursts
appear to be short: the average profile of the bursts
(Fig.~\ref{figavburst}) has an e-folding decay time of $6\pm2$~s and
shows no flux after about 15 s. Due to the limits of the WFC
sensitivity, we are unable to study these bursts in any spectral
detail, except for the brightest burst in a rather limited sense, and
we cannot confirm the type-I classification. Furthermore, we cannot
claim to have a complete database of bursts. Our observations suggest
that the source bursts regularly except possibly during late 1997
through early 1998, but the gap may be due to pure chance. The average
observed burst rate is 2.5$\pm0.5$ per Ms. Thus, one would expect on
average 2.1 bursts during the gap.  Based on Poissonian statistics,
the probability to find none by coincidence is 12\%.

\begin{figure}[t]
\psfig{figure=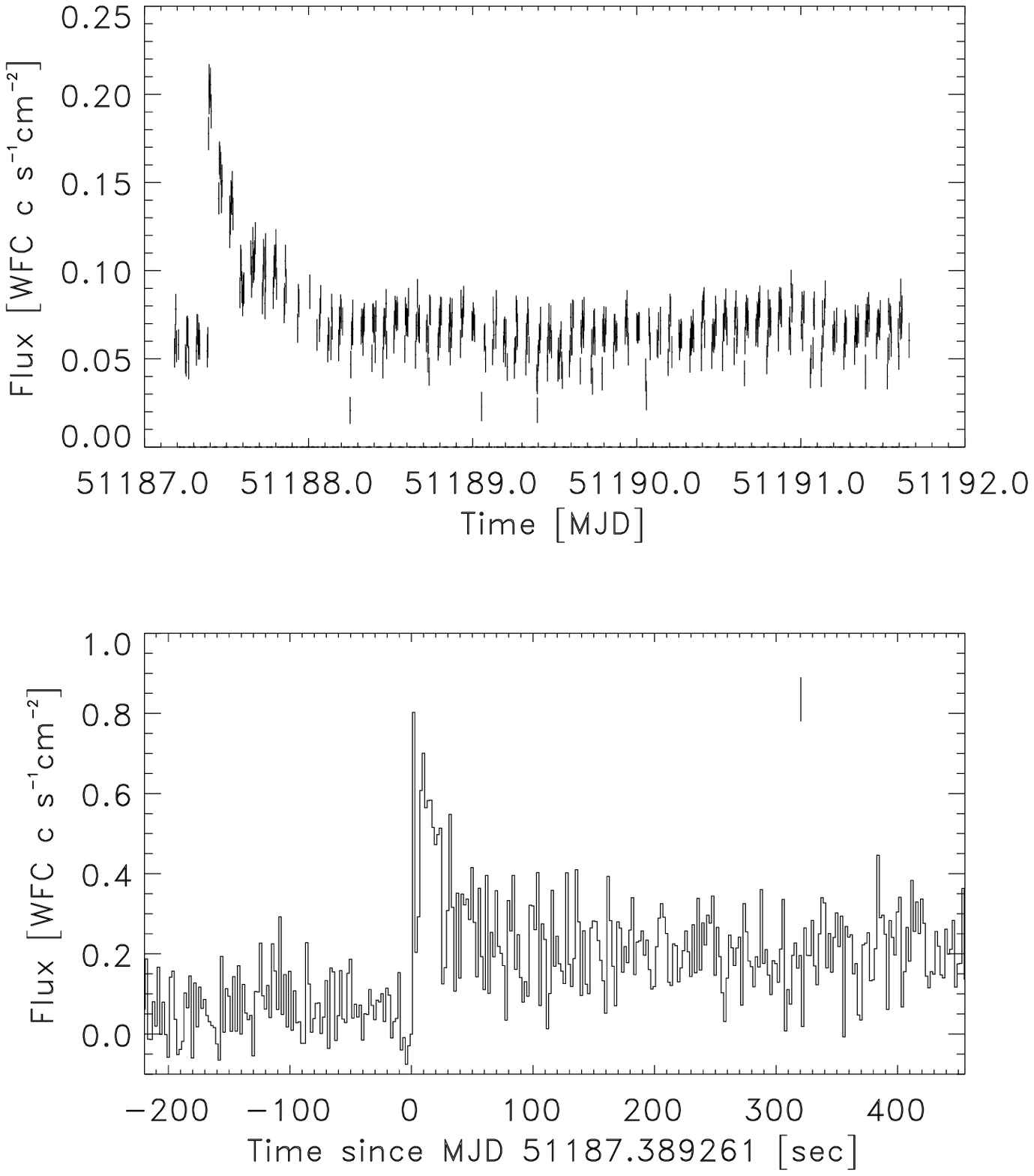,width=\columnwidth,clip=t}
\caption{Light curves obtained with the WFC on January 9-13, 1999 at
300~s resolution (top panel) and focused on the burst onset at 2~s
resolution (bottom panel, with vertical line indicating a
typical 1$\sigma$ error). The bandpass is 2--28 keV.
\label{figwfc1}}
\end{figure}

In Fig.~\ref{figwfc1} we present the light curve as measured with WFC
unit 1 during a long observation between January 9.2 and 13.7, 1999,
when a flare occurred reminiscent of a superburst. This is the only
such flare we found in the whole data set for \bron.  The flare onset
is on 9.389 January (UT), when a short normal X-ray burst
occurred. This ordinary X-ray burst is the brightest of the sample of
26. It appears to consist of two peaks roughly 8~s apart. The question
arises whether the drop in flux is due to dipping activity in this
established dipper. During the observation clear dipping activity is
observed several times, for instance at MJD~51187.597, 51188.255 and
51189.072.  From the 3.93336~hr orbital period (Motch et al. 1987) we
predict that the period of dipping activity nearest to the burst onset
is between MJD~51187.416 and 51187.450. This interval starts 2310~s
after the burst onset and is during a data gap.  Thus, the
double-peaked nature of the burst unlikely is related to dipping and
is probably intrinsic to the neutron star. The flare following the
normal X-ray burst lasts approximately 14~hr. It decays in an
exponential-like manner with a fitted e-folding time of $6.0\pm0.3$~hr
(after excluding the dipping intervals; $\chi^2_\nu=1.47$ with 488
dof). During the five days following the superburst the decay levels
off to a flux level that is 10$\pm$3\% higher than the pre-burst
level.

We do not find any evidence for the flare to start prior to the normal
burst, like for the second superburst in 4U~1636--536 (Strohmayer \&
Markwardt 2002), but the statistical quality of our data is rather
limited to make a conclusive statement. In a two-minute interval prior
to the normal burst, the highest flux consistent with our data bridges
20\% of the difference between the pre-burst flux and the superburst
peak flux (excluding the normal burst). This is roughly equal to what
was measured in 4U~1636-536 (Strohmayer \& Markwardt 2002).

\section{Confirmation of superburst nature}

\begin{table}
\caption[]{Characteristics of \bron\ superburst.  A distance of 13~kpc
has been assumed (see text).  For comparison the range of values for
the other seven superbursts has been provided in the last column if
useful (in the same units; from Kuulkers et al. 2002 and Strohmayer \&
Markwardt 2002).\label{tab1}}
\begin{tabular}{lll}
\hline\hline
Duration  & 14$\pm2$ hr & 4--12 \\
Precursor? & Yes & 3 times \\
$\tau_{\rm exp}$ (hr) & $6.0\pm0.3$ & 1--2 \\
k$T_{\rm max}$ (keV) & $1.8\pm0.1$ & 2--3.0 \\
$L_{\rm peak}$ (10$^{38}$\lum) & $0.44\pm0.2$ & 0.8--3.4 \\
$L_{\rm pers}$ ($L_{\rm edd}$) & 0.13$\pm0.03$ & 0.1--0.25 \\
$E_{\rm b}$ (10$^{42}$~erg) & $0.8\pm0.2$ & 0.5~--~$>1.4$ \\
$\tau$ (=$E_{\rm b}/L_{\rm peak}$) (hr) & $5.0\pm1.7$ & 1.2--2.0 \\
$\gamma$ (=$L_{\rm pers}/L_{\rm peak}$) & $0.66\pm0.16$ & 0.1--0.5 \\
$t_{\rm no\hspace{1mm}bursts}$ (d) & $<$124.7 & $>7$ \\
Donor & H/He$^\ddag$ & He and H/He \\
\hline\hline
\end{tabular}

\noindent
$^\ddag$From Motch et al. 1987.
\end{table}

\begin{figure}[t]
\psfig{figure=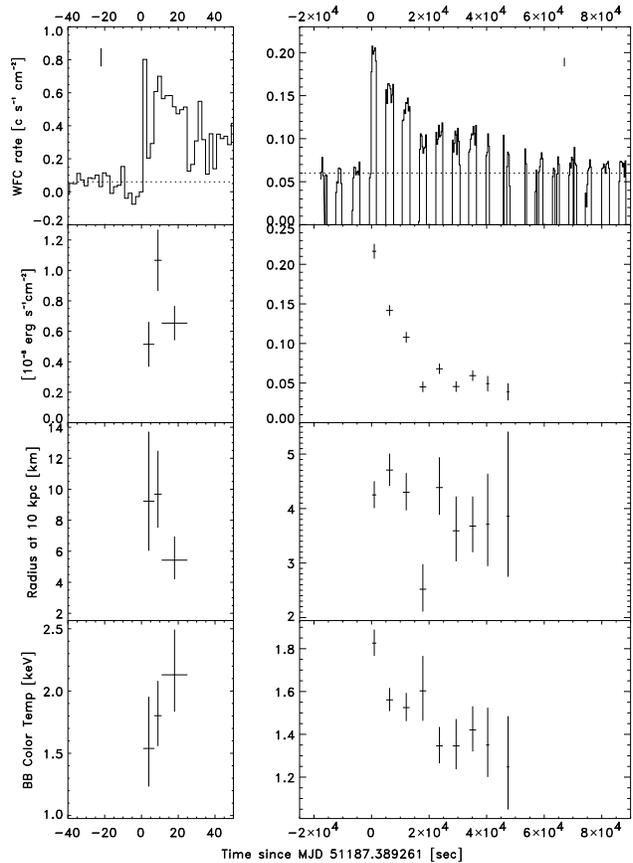,width=\columnwidth,clip=t}
\caption{Time-resolved spectroscopy of the superburst (right) and the
normal type-I precursor burst (left). Note the different y scales in
both columns of plots. In the top panels, a dashed line indicates the
pre-burst flux level for guiding purposes and solid vertical lines
indicate the typical 1$\sigma$ error. The time resolution of the
observed photon flux is 2~s in the left and 300~s in the right panel.
In the other panels errors are 1$\sigma$ values.  The 2nd row of
panels show the bolometric flux of the black body component. The
decreased flux levels near $1.8\times10^4$ and $7.5\times10^4$~s are
due to dipping activity.
\label{figwfc2}}
\end{figure}

Type-I X-ray bursts, including superbursts, are characterized by black
body continua and cooling during the decay. Therefore, two crucial
tests are that the black body radiation model is a better description
than any other reasonable model and that the temperature decreases.
To study the possibly varying spectrum during the flare, we generated
ten spectra: one for each of nine 100-min long BeppoSAX orbits during
the flare and one for all pre-burst data combined. The resulting
spectra have sufficient statistical quality for a meaningful analysis.
First, we modeled these data with absorbed thermal bremsstrahlung.  We
kept the hydrogen column density that parameterizes the low-energy
interstellar absorption (Morrison \& McCammon 1983) fixed to
3.2$\times10^{21}$~cm$^{-2}$ (see Iaria et al. 2001 and Boirin \&
Parmar 2003) and note that the 2--28 keV spectrum is not very
sensitive to such low densities.  While the thermal bremsstrahlung
model describes the pre-burst data well ($\chi^2_\nu=0.65$), it does
not do so for the flare data ($\chi^2_\nu=2.46$, 2.02, 1.86 and 1.31
for the first 4 flare spectra with $\nu=26$). Adding a black body
component while tying the bremsstrahlung parameter values to those for
the pre-burst spectrum significantly improves the fit
($\chi^2_\nu=0.69$, 0.67, 0.82 and 1.01 for the first 4 flare spectra
with $\nu=26$). Substituting the black body model for a second
bremsstrahlung model does not provide an improvement
($\chi^2_\nu=3.08$, 2.40, 2.15 and 1.57 for the first 4 flare spectra
with $\nu=26$) and neither does substitution by a power law
($\chi^2_\nu=3.72$, 3.18, 2.63 and 1.69 for the first 4 flare spectra
with $\nu=26$). We conclude that a black body model describes the
flare spectrum best. The fit results are given in
Fig.~\ref{figwfc2}. It is clear that the black body temperature
decreases during the decay of the flare. Thus, the flare has all the
characteristics of other superbursts and we confirm the superburst
nature.  In Table~\ref{tab1} we list the basic characteristics of this
superburst.

Also plotted in Fig.~\ref{figwfc2} is the time-resolved spectroscopy
of the precursor burst. This is the brightest normal burst that we
detected.  Although the black body model is an adequate description of
the data, and the resulting temperature typical for type-I X-ray
bursts, we formally cannot confirm this as an X-ray burst because we
cannot prove cooling.  Nevertheless, \bron\ is a known burster with
bursts of similar duration and we consider it very unlikely that this is
something else than a type-I burst.

The precursor burst has a remarkable feature: it is double peaked.
This suggests that the burst invoked strong photospheric radius
expansion due to near-Eddington luminosities (e.g., Lewin et
al. 1993). This would be the first detection of a radius-expansion
burst in \bron. Keeping in mind that the statistical quality of our
data is insufficient to confirm the expansion directly through black
body model fits, we derive a distance estimate: $d=13\pm3$~kpc. This
is consistent with previous estimates (Courvoisier et al. 1986 and
Motch et al. 1987).

We note that the 2--28 keV spectrum after the superburst (i.e., after
MJD~51188.0; see Fig.~\ref{figwfc1}) can be adequately modeled by
thermal bremsstrahlung. The temperature varies between $5.2\pm0.2$ and
6.5$\pm0.3$ keV and the 2--28 keV flux varies between 7.7 and
9.2$\times10^{-10}$~\ecs\ on a 1-day time basis (6.7 to
7.3$\times10^{-10}$~\ecs\ between 2 and 10 keV). 

\section{Discussion}
\label{discuss}

The fortunate circumstance exists that a high-quality broad-band
spectrum was taken of \bron\ just two weeks before the
superburst. Iaria et al.  (2001) determined the 2--10 keV flux at
6.9$\times10^{-10}$~\ecs\ which compares well to the flux in the days
after the superburst. The unabsorbed 0.1 to 100 keV flux on 23
December 1998 was 1.4$\times10^{-9}$~\ecs. For a distance of 13~kpc
this translates to a luminosity of 2.9$\times10^{37}$~\lum. This is
roughly 15\% of the Eddington limit for a 1.4~M$_\odot$ neutron star
with a hydrogen-dominated atmosphere (2$\times10^{38}$~\lum). Thus,
the superburst follows the rule of the other superbursts in occurring
on objects that accrete at 5 to 10 times below Eddington (Wijnands
2001; Kuulkers et al. 2002).

The suggestion is strong in three previous cases that a superburst
quenches normal type-I bursting activity for tens of days (Cornelisse
et al. 2000, 2002, Kuulkers et al. 2002).  For \bron\ this is hard to
prove, given the low burst frequency.  The last normal burst prior to
the superburst was seen 51 days earlier, and the first one following
the superburst 125 days later (while it should be noted that there is
a long data gap from 20 to 120 days after the superburst). Therefore,
our data do not prove nor disprove the presence of quenching in \bron.

The superburst from \bron\ shows extremes in two parameters
(Table~\ref{tab1}): the peak luminosity is two times lower than the
previous minimum value and the decay time is three times
longer than the previous maximum.  These two extremes approximately
cancel out in the total burst energetics. The long decay must be due
to either a relatively deep flash location, a smaller neutron star
mass, the presence of relatively heavy elements, or a combination
of these circumstances (Cumming \& Bildsten 2001).

The superburst from \bron\ is the fourth for which coincidence with a
normal type-I burst could be proven. Like in 4U 1636-536 (Strohmayer
\& Markwardt 2002) and KS 1731-260 (Kuulkers et al. 2002), the
precursor from \bron\ has a peak flux that is higher than that of the
superburst (roughly three times).  This emphasizes the exceptional
superburst from 4U~1820-30 (Strohmayer \& Brown 2002) with a precursor
that was {\em weaker\/} than the superburst which must be related to
the exceptional nature of the donor star.

In Table~\ref{tab2} we list a few burst properties for all
superbursters and for six bursters which have not yet shown
superbursts. The properties are averaged over the complete WFC
database.  One is the $\alpha$-parameter which is defined as the ratio
of the persistent fluence between bursts and the burst fluence. Since
there has not yet been a systematic spectral analysis of the thousands
of WFC-detected bursts, we employ an alternative $\alpha$ which is
based on the ratio of the {\em number of observed photons} rather than
on {\em intrinsic radiation energy}. This is a useful alternative
since we only compare data from a single instrument. For a few
sources, verification with the literature of the alternative $\alpha$
with the true $\alpha$ (see Table~\ref{tab2}) shows that the two are
not far apart. It turns out that superbursters generally have much
higher average $\alpha$-values\footnote{$\alpha$ may occasionally be
low while not significantly affecting the high average over long time
scales.  This is particularly well documented for 4U~1820-303 (e.g.,
Cornelisse et al. 2003).} than the other bursters.

Another property is the e-folding decay time of the average 2--28 keV
burst profile. We generated such profiles by selecting bursts with
peak significances in excess of 3$\sigma$ at 1~sec resolution,
aligning the bursts at their peaks, and averaging the fluxes weighted
by the inverse square of the statistical $1\sigma$ error.  An
exponential function was fitted to the tail of the average profile
(from 1~sec on after the peak) with as free parameters the
normalization, e-folding decay time and background level. It turns out
that superbursters have generally shorter decay times than the other
bursters, with a boundary value of about 6~s. A decay time as short as
6~s indicates the presence of a helium-rich layer. Such a layer is
formed through stable burning of accreted hydrogen at global mass
accretion rates in excess of one tenth the Eddington limit (Fujimoto
et al. 1981; Bildsten 1998) as is the case for the superbursters,
except for 4U~1820-303 which is thought to have a helium-rich donor
star (for a recent discussion on that, see Cumming 2003).

The high $\alpha$ suggests that a sizeable fraction of the helium is
burned in a {\em stable manner} rather than an unstable burst-like
manner (e.g., Van Paradijs et al. 1988a).  Thus, favorable conditions
for a superburst appear to include the presence of a pure helium layer
which is burned in a stable as well as unstable fashion. This supports
inferences by Strohmayer \& Brown (2002), Cumming (2003) and Woosley
et al. (2003) that only stable helium burning can generate sufficient
amounts of carbon for triggering a superburst.

We note that EXO~0748-676, 4U~1728-34 (GX~354-0) and GS~1826-24 are
examples of frequently bursting sources that do not show high $\alpha$
values. These sources will, therefore, unlikely exhibit
superbursts. Conversely, our analysis identifies one new prospective
superburster: 4U~1705-44.

\begin{table}
\caption[]{Average burst properties of all superbursters (above the
dividing line) and six non-superbursters, as observed with
BeppoSAX-WFC.\label{tab2}}
\begin{tabular}{llrrc}
\hline\hline
Object name  & $T_{\rm C}^{\rm (a)}$ & $\alpha^{\rm (b)}$ & $\alpha^{\rm (c)}$ & $\tau^{\rm (d)}$ [sec]\\
\hline
4U~1254-690  & 4.6         & 4800 &                     & $6\pm2$ (15) \\ 
4U~1636-536  & 0.6         &  440 & 44-336$^{\rm [1]}$    & $6.2\pm0.1$ (67) \\
KS~1731-260$^{\rm (e)}$ & 0.8&780 & 30-690$^{\rm [2]}$    & $5.6\pm0.2$ (37) \\
4U~1735-444  & 2.4         & 4400 & 220-7728$^{\rm [3]}$  & $3.2\pm0.3$ (34) \\
GX~3+1       & 1.2         & 2100 & 1700-               & \\
             &             &      & 21000$^{\rm [4]}$     & $4.6\pm0.1$ (61) \\
4U~1820-303  & 1.5         & 2200 &                     & $4.5\pm0.2$ (47) \\
Ser X-1      & 2.9         & 5800 &                     & $5.7\pm0.9$ (7) \\
\hline						    
EXO~0748-676 & 1.0         &  140 & 18-34$^{\rm [5]}$     & $12.8\pm0.4$ (155) \\
4U~1702-429  & 0.3         &   58 &                     & $7.7\pm0.2$ (107) \\
4U~1705-44   & 1.1         & 1600 & 55--1455$^{\rm [6]}$  & $8.7\pm0.4$ (74) \\
GX~354--0    & 0.2         &   97 & 105-140$^{\rm [7]}$   & $4.7\pm0.1$ (417) \\
A~1742-294   & 0.4         &  130 &                     & $16.8\pm1.0$ (141) \\
GS~1826-24   & 0.2         &   32 & 41$^{\rm [8]}$        & $30.8\pm1.5$ (248) \\
\hline\hline
\end{tabular}

\noindent
$^{\rm (a)}$ average wait time between bursts in days; $^{\rm
(b)}\alpha$ is ratio of average persistent 2--28 keV flux (in WFC
c~s$^{-1}$cm$^{-2}$) times average wait time between two bursts (2nd
column) and burst fluence (in WFC c~cm$^{-2}$); $^{\rm (c)}\alpha$
from literature if covering the largest persistent flux range known
($^{[1]}$=Lewin et al. 1987; $^{[2]}$=Muno et al. 2000, $^{[3]}$=van
Paradijs et al. 1988b; $^{[4]}$=den Hartog et al.  2003;
$^{[5]}$=Gottwald et al. 1987; $^{[6]}$=Gottwald et al. 1989;
$^{[7]}$=Basinska et al. 1984; $^{[8]}$=Galloway et al. 2003); $^{\rm
(d)}$e-folding decay time of the average 2--28 keV burst profile;
between parentheses the number of bursts averaged; $^{\rm (e)}$This is
a transient and only data are given for persistent flux levels
comparable to when the superburst occurred.

\end{table}

The WFC observations have thus far resulted in the detection of four
superbursts. The effective observation time for superburst
detection\footnote{this is simply the elapse time of an observation,
including the data gaps due to earth occultations and passes
over the South Atlantic Anomaly which are shorter than half an hour.}
added over all 27 persistently bright X-ray bursters is 7.9~yr (in
other words, the average observation time per source is 0.3~yr). Thus,
the average superburst frequency is $0.51\pm0.25$ per year per
object. Considering that superbursts may occur in only a selection of
X-ray bursters (like those with persistent flux levels in excess of
0.1 times Eddington), the rate may actually be larger by up to a
factor of two. Therefore, the WFC observations yield an average
superburst recurrence time that lies between 0.7 and 4 years (at 70\%
confidence).

\acknowledgement 

JZ acknowledges support from the Netherlands Organization for
Scientific Research (NWO). Gerrit Wiersma, Jaap Schuurmans, Nuovo
Telespazio and the ASI BeppoSAX Science Data Center are acknowledged
for continued support, and the ASM/RXTE team for providing ASM
standard data products to the public.

\end{document}